\documentclass[aps,twocolumn,pra,showpacs,floatfix]{revtex4}
\usepackage{dcolumn}
\usepackage{amsmath}
\usepackage{multirow}
\usepackage{booktabs}

\begin{document}

\title{\bf Optical atomic clocks with suppressed black body radiation  shift.}
\author{A. Kozlov, V. A. Dzuba, and V. V. Flambaum}
\affiliation{School of Physics, University of New South Wales, 
Sydney 2052, Australia}
\date{\today}

\begin{abstract}
We study a wide range of neutral atoms and ions suitable for
ultra-precise atomic optical clocks with naturally suppressed black
body radiation shift of clock transition frequency. Calculations show
that scalar polarizabilities of clock states cancel each other for at
least one order of magnitude for considered systems. Results for
calculations of frequencies, quadrupole moments of the states, clock
transition amplitudes and natural widths of upper clock states are
presented.   
\end{abstract} 
\pacs{PACS: }
\maketitle

\section{Introduction}

\begin{table}\center
\caption{BBR shift at room temperature of existing and prospective atomic clocks. If available, uncertainties are given in parenthesis.}
{\renewcommand{\arraystretch}{0}%
\begin{tabular}{l c l l l}
\hline\hline
\rule{0pt}{4pt}\\
Z & element & transition & $\beta_{BBR}, \times 10^{-18}$ & reference\\ 
\rule{0pt}{4pt}\\
\hline
\rule{0pt}{2pt}\\
13 & Al$^{+}$ & $^1\text{S}_0\rightarrow ^3\text{P}_0$& 3.8(0.4) & \cite{Safronova:2012}\\
\rule{0pt}{2pt}\\
\hline
\rule{0pt}{2pt}\\
38 & Sr$^{+}$ & $^2\text{S}_{1/2}\rightarrow ^2\text{D}_{5/2}$ & 670(250) & \cite{Madej:2004}\\
\rule{0pt}{2pt}\\
\hline
\rule{0pt}{2pt}\\
38 & Sr & $^1\text{S}_0\rightarrow ^3\text{P}_0$& 5500(70) & \cite{Porsev:2006}\\
\rule{0pt}{2pt}\\
\hline
\rule{0pt}{2pt}\\
40 & Zr$^{2+}$ & $^3\text{F}_2\rightarrow ^3\text{P}_0$& 9& this work\\
\rule{0pt}{2pt}\\
\hline
\rule{0pt}{2pt}\\
40 & Zr & $^3\text{F}_2\rightarrow ^3\text{P}_0$& 621& this work\\
\rule{0pt}{2pt}\\
\hline
\rule{0pt}{4pt}\\
47 & Ag & $^2\text{S}_{1/2}\rightarrow ^2\text{D}_{5/2}$& 190 & \cite{Topcu:2006}\\
\rule{0pt}{2pt}\\
\hline
\rule{0pt}{4pt}\\
52 & Te & $^3\text{P}_2\rightarrow ^3\text{P}_0$& 112& this work\\
\rule{0pt}{2pt}\\
\hline
\rule{0pt}{2pt}\\
53 & I$^+$ & $^3\text{P}_2\rightarrow ^3\text{P}_0$& 15& this work\\
\rule{0pt}{2pt}\\
\hline
\rule{0pt}{2pt}\\
54 & Xe$^{2+}$ & $^3\text{P}_2\rightarrow ^3\text{P}_0$& 4& this work\\
\rule{0pt}{2pt}\\
\hline
\rule{0pt}{2pt}\\
68 & Er$^{2+}$ & $^3\text{H}_6\rightarrow ^3\text{F}_4$& $<$63& this work\\
\rule{0pt}{2pt}\\
\hline
\rule{0pt}{2pt}\\
68 & Er & $^3\text{H}_6\rightarrow ^3\text{F}_4$&  $<$570 & this work\\
\rule{0pt}{2pt}\\
\hline
\rule{0pt}{2pt}\\
69 & Tm$^{3+}$ & $^3\text{H}_6\rightarrow ^3\text{F}_4$& $<$3& this work\\
\rule{0pt}{2pt}\\
\hline
\rule{0pt}{2pt}\\
70 & Yb$^{+}$ & $^2\text{S}_{1/2}\rightarrow ^2\text{D}_{3/2}$ & 580(30) & \cite{Schneider:2005}\\
\rule{0pt}{2pt}\\
\hline
\rule{0pt}{2pt}\\
70 & Yb$^{+}$ & $^2\text{S}_{1/2}\rightarrow ^2\text{F}_{7/2}$ & 234(110) & \cite{Lea:2006}\\
\rule{0pt}{2pt}\\
\hline
\rule{0pt}{2pt}\\
70 & Yb & $^1\text{S}_0\rightarrow ^3\text{P}_0$& 2400(250) & \cite{Porsev:2006}\\
\rule{0pt}{2pt}\\
\hline
\rule{0pt}{2pt}\\
71 & Lu$^{+}$ & $^1\text{S}_0\rightarrow ^3\text{D}_1$& 54& this work\\
\rule{0pt}{2pt}\\
\hline
\rule{0pt}{2pt}\\
72 & Hf & $^3\text{F}_2\rightarrow ^3\text{P}_0$& 855& this work\\
\rule{0pt}{2pt}\\
\hline
\rule{0pt}{2pt}\\
84 & Po & $^3\text{P}_2\rightarrow ^3\text{P}_0$& 185& this work\\
\rule{0pt}{2pt}\\
\hline
\rule{0pt}{2pt}\\
90 & Th & $^3\text{F}_2\rightarrow ^3\text{P}_0$& 303& this work\\
\rule{0pt}{2pt}\\
\hline
\rule{0pt}{2pt}\\
91 & Pa$^{3+}$ & $^3\text{H}_4\rightarrow ^3\text{F}_2$& 21& this work\\
\rule{0pt}{2pt}\\
\hline
\rule{0pt}{2pt}\\
91 & Pa$^{3+}$ & $^3\text{F}_2\rightarrow ^3\text{P}_0$& 20& this work\\
\rule{0pt}{2pt}\\
\hline\hline
\end{tabular}}\label{Tab:1}

\end{table}

Atomic clocks are one of the most accurate tools ever designed that
found application in many different modern technologies. Both optical
lattice atomic clock \cite{Bloom:2014} and ion clock \cite{Chou:2010}
have demonstrated fractional accuracy at the level of few parts times
$10^{-18}$. There is a number of systematic shifts that one needs to
overcome at this level of accuracy. One of the most challenging from
technical point of view is the black body radiation (BBR) shift. The
typical ways of removing this shift are either colling entire device
to cryogenic temperatures \cite{Bize:2003} or building a sophisticated
thermal shields that allow to stabilize BBR shift, measure it and
subtract later \cite{Bloom:2014}. Both methods lead to considerable
increase of complexity, size and price of the device.  

It was suggested in \cite{Berengut:2010,Berengut:2011,Berengut:2012,Dzuba:2012} to use highly charged ions (HCI)
for atomic clock purposes. Apart from many other advantages HCI have
naturally suppressed BBR shift due to small values of scalar
polarizabilities of clock states. Authors of \cite{Dzuba:2012}
demonstrated that HCI with $nl^2$ two-electrons or two-holes
configuration have optical transitions within the same configuration
that allows to use them as an atomic clocks. In these systems there
are ground and long-living first excited states with allowed electric
quadrupole transition withing optical or infrared frequency range. It
was pointed that the width of first excited state of the $nf^{12}$
two-hole configuration was estimated to be an order of magnitude
smaller than the one of two-electron $nf^2$ first excited state with
the same transition frequency. An important conclusion of
\cite{Dzuba:2012} is that clock transitions in HCI have many orders of
magnitude larger quality factors than the ones found in modern atomic
clocks. In the same time handling HCI is much more sophisticated task
compared to working with low charged ions (LCI) and neutral atoms. 

In this paper we investigate another way of reducing BBR shift in
atomic clocks. The proposed systems are neutral atoms and low charged
ions (LCI). We show that selected elements possess large quality
factors and small BBR shifts. In the same time manipulations with them
are accessible with developed experimental methods. We consider
systems with suitable electric quadrupole clock transition within the
same configuration. For LCI accurate numerical calculations show that
for two-electron and two-hole configurations one can write the
following inequality for transition matrix element values between
ground and first excited states of the same configuration
$A\left(nf^{12}\right) < A\left(nf^2\right) < A\left((n+1)d^8\right) <
A\left((n+1)d^2\right) < A\left((n+2)p^4\right)$. Every next
configuration in this sequence has about half an order of magnitude
larger transition matrix element compared to the previous one. In the
same time in our recent paper \cite{Er} we have investigated the $4f^{12}$
configuration for doubly ionized erbium together with the $4f^{12}6s^2$ of
the neutral one. Since the $6s^2$ electrons form closed shell, energy
level structure of these configurations is almost identical for both
Er I and Er III. Clock transitions in both neutral and doubly ionized
erbium have almost the same frequency that differs only by 46
cm$^{-1}$. The radiation widths of corresponding excited state in
neutral erbium were only several times larger than the one of doubly
ionized erbium. We anticipate the same inequality to hold for
two-electron or two-hole neutral atoms with extra $(n+2)s^2$
electrons.  

Important disadvantage of proposed elements is large value of total
angular momentum $J$ in one or both clock states. When $J \ge 1$ the
atom or ion has a non-zero quadrupole momentum
which couples to the electric field gradient. It can be
especially important for optical lattice atomic clocks since the electric
field gradient of trap laser can have relatively high values. In
order to estimate this effect we perform calculations of quadrupole
moments for considered systems.  

\section{Systematic shifts in atomic clocks}

A number of systematic shifts affect and limit the accuracy of atomic
clocks. Among the main ones there are black body radiation shift
(BBR), interaction of atomic quadrupole moments with gradients of
electric field, micro and secular motion, Stark and Zeeman shifts,
background-gas collisions, gravitational shift, etc. Some of these
factors were discussed in \cite{Chou:2010, Derevianko:2012}. The most
significant factors are BBR, quadrupole and Zeeman shifts. Zeeman
shift and other effects due to influence of the external magnetic field on
the clock transition were widely investigated (see for example
\cite{Chou:2010, Rosenband:2007, Bernard:1998}), well known methods
are developed in order to minimize or cancel corresponding shifts. In
the same time the black body radiation shift (BBR) remains the most
significant obstacle on the way to more accurate and compact atomic
clocks. As was mentioned in introduction for proposed elements
quadrupole shift may also be essential and requires consideration. 

\subsection{Black Body Radiation shift}
The BBR shift originates from perturbation of the clock states by the
environment photon bath due to dynamic Stark shift. The magnitude of
this shift is given by the following equation \cite{Porsev:2006} 
\begin{equation}\label{BBR} 
\frac{\Delta\omega}{\omega_0}\left|_{\text{\parbox{.3in}{BBR}}}\approx
  -\frac{2\pi^3\alpha^3}{15}\frac{T^4}{\omega_0}\Delta\alpha_0\right. \equiv
\beta_{BBR}\left(\frac{T}{300K}\right)^4, 
\end{equation}
where $T$ is the temperature, $\alpha$ is the fine structure constant,
$\omega_0$ is the unperturbed clock transition frequency, $\Delta\alpha_0$
is the difference of scalar polarizabilities of the clock states,
$\Delta\alpha_0=\alpha_0(e)-\alpha_0(g)$.
The values for $\beta_{BBR}$ for some known clocks as well as the ones
investigated in this paper are listed in table \ref{Tab:1}. 

Scalar polarizability $\alpha_0(a)$ can be expressed via sums over complete sets of intermediate states involving matrix elements of the electric dipole operator $\mathbf{D}$ (in coordinate representation $\mathbf{D} = -e\sum_i \mathbf{r}_i=\sum_i \mathbf{d}_i$)
\begin{equation}
\alpha_0(a) = \frac{2}{3(2J_a+1)} \sum_n \frac{\langle a||\mathbf{D}|| n
  \rangle^2}{E_a - E_n}
\label{alpha}.
\end{equation}
Here $|a\rangle$ and $|n\rangle$ are many-electron atomic states and
$E_a$ and $E_n$ are corresponding energies. 

Currently the best atomic clocks have fractional accuracy level of
$\Delta \omega/\omega_0 \sim 7\times10^{-18}$ for aluminum ion clock
\cite{Chou:2010} and $\Delta \omega/\omega_0 \sim 6.4\times10^{-18}$
for optical lattice strontium clocks \cite{Bloom:2014}. Aluminum ion
clock is the only clock (operating at room temperature) where
fractional BBR shift is under $10^{-17}$ level due to almost 98\%
cancellation of the clock state scalar
polarizabilities~\cite{Safronova:2012}. The rest of 
the clocks require either separate measurement of BBR shift and
further thermal stabilization \cite{Bloom:2014, Hinkley:2013, Le
  Targat:2013} or cooling to cryogenic temperatures \cite{Bize:2003}. 

\subsection{Quadrupole shift}\label{QS}
Coupling of the external electric field gradient to an atomic quadrupole
moment leads to the emergence of a significant systematic shift. If the electric
field is aligned along quantization axis, the corresponding term in
atomic Hamiltonian can be written as 
 
\begin{equation}\label{quadr}   
H_Q=\frac{1}{2}Q_a\frac{\partial E_z}{\partial z},
\end{equation}
where $Q_a$ is the quadrupole moment of atom, given by the following equation
\begin{eqnarray}\label{quad}   
&&Q_a=2\left<J_aJ_a|E2|J_aJ_a\right>=\nonumber\\
&&\left<J_a||E2||J_a\right>\sqrt{\frac{J_a(2J_a-1)}{(2J_a+3)(2J_a+1)(J_a+1)}},
\end{eqnarray}
where $J_a$ is the total electron angular momentum, $\left<a||E2||a\right>$
is the reduced matrix element of electric quadrupole transition
operator. Using (\ref{quadr},\ref{quad}) one can obtain the following
expression for the frequency shift between two clock states: 
\begin{equation}\label{omega0}
\omega=\omega_0+(C_{J_g,M_g}Q_{g}-C_{J_e, M_e}Q_{e})\frac{\partial
  E_z}{\partial z}, 
\end{equation}
where $\omega_0$ is unperturbed transition frequency, $Q_g$ and $Q_e$
are ground and excited states quadrupole moments respectively,
coefficients $C_{J,M}$ depend on the projection $M$ of the total
angular momentum $J$:
\begin{equation}
C_{J,M}=\frac{3M^2-J(J+1)}{3J^2-J(J+1)}.
\end{equation}
Estimates for the magnitude of relative quadrupole shift in neutral
and ionized erbium can be found in \cite{Er}. The values of typical
electric field gradients in ion trap $\partial E_z/\partial z \sim
10^{6} \ {\rm V/m}^2$~\cite{Barwood:2004} that leads to the relative frequency
shift for double ionized erbium is $\Delta \omega_Q/\omega_0 \sim
10^{-15}$, while for optical lattice clocks on neutral erbium these
values are $\partial E_z/\partial z \sim 10^{7} \ {\rm
  V/m}^2$~\cite{Porsev:2004} and $\Delta \omega_Q/\omega_0 \sim 10^{-14}$
respectively. For other atoms and ions the relative quadrupole shifts
may be significantly larger 
and therefore require accurate treatment. There are several ways of
suppression or cancellation quadrupole shift in atomic clocks. They
were considered in details in \cite{Itano:2000, Dube:2005, Roos:2006,
  Er} and allow to achieve several orders of magnitude cancellation of
quadrupole shift. 

It should be pointed that if total angular momentum of an atom $F=0, 1/2$
or total electronic angular momentum $J=0, 1/2$ then the quadrupole
momentum of corresponding state is equal to zero. Therefore it becomes
sometimes possible to cancel the quadrupole shift if any of the latter
conditions holds for both clock states. For most of considered
elements listed in table \ref{Tab:1} upper clock state has $J=0$,
therefore quadrupole shift for this states vanishes.  
\begin{table*}\center
\caption{Scalar polarizabilities for different levels of ground state
  configuration for tin and doubly ionized zirconium. Experimental
  data is taken from NIST atomic spectra database. Units for energy
  are inversed centimeters, polarizabilities are given in atomic
  units.} 
{\renewcommand{\arraystretch}{0}%
\begin{tabular}{l c c c r r c r c}
\hline\hline
\rule{0pt}{4pt}\\
Z & element & config. & term & experimental & \multicolumn{2}{c}{calc. relativistic} & \multicolumn{2}{c}{calc. non-relativistic} \\ 
\rule{0pt}{2pt}\\
& & & & energy, cm$^{-1}$ & energy, cm$^{-1}$ & polarizability & energy, cm$^{-1}$ & polarizability \\ 
\rule{0pt}{4pt}\\
\hline
\rule{0pt}{4pt}\\
& & & $^3P_0$ & 0 & 0 & 50.5 & 0 & 54.7\\
\rule{0pt}{2pt}\\
 &  &  & $^3P_1$ & 1691.81 & 1720.74 & 52.3 & 0.33 & 54.7\\
\rule{0pt}{2pt}\\
50& Sn & $5p^2$ & $^3P_2$ & 3427.67 & 3584.07 & 53.7 & 3.12 & 54.7\\
\rule{0pt}{2pt}\\
& & & $^1D_2$ & 8612.96 & 9536.59 & 58.3 & 6087.3 & 58.5\\
\rule{0pt}{2pt}\\
& & & $^1S_0$ & 17162.50 & 18396.06 & 65.4 & 14680.56 & 64.5\\
\rule{0pt}{2pt}\\
\hline
\rule{0pt}{2pt}\\
& & & $^3F_2$ & 0 & 0 & 11.1 & 0 & 10.4\\
\rule{0pt}{2pt}\\
& & & $^3F_3$ & 681.59 & 729.55 & 11.1 & -2.81 & 10.4\\
\rule{0pt}{2pt}\\
& & & $^3F_4$ & 1486.45 & 1061.73 & 11.1 & -6.48 & 10.4\\
\rule{0pt}{2pt}\\
40 & Zr$^{2+}$ & $4d^2$ & $^1D_2$ & 5743.39 & 6601.21 & 13.6 & 6434.48 & 12.3\\
\rule{0pt}{2pt}\\
& & & $^3P_0$ & 8063.63 & 8223.31 & 11.3 & 7858.93 & 10.7\\
\rule{0pt}{2pt}\\
& & & $^3P_1$ & 8327.12 & 8504.47 & 11.3 & 7856.38 & 10.7\\
\rule{0pt}{2pt}\\
& & & $^3P_2$ & 8839.97 & 9097.43& 11.5 & 7853.86 & 10.7\\
\rule{0pt}{4pt}\\
\hline\hline
\end{tabular}}\label{Tab:4}
\end{table*}

\begin{table}\center
\caption{Scalar polarizabilities of ground (J=6) and first excited (J=4) states for highly charged ions \cite{Dzuba:2012} sequence with configuration $\left[\text{Pd}\right]5s^24f^{12}$ for Hf$^{12+}$ and W$^{14+}$ and $\left[\text{Pd}\right]4f^{12}$ for rest of the atoms. $\alpha_0(g)$ and $\alpha_0(e)$ are the scalar polarizabilities of ground and first exited states respectively, their values are given in $a_0^3$.}
{\renewcommand{\arraystretch}{0}%
\begin{tabular}{l c l l l}
\hline\hline
\rule{0pt}{4pt}\\
Z & element & $\alpha_0(g)$, a.u. & $\alpha_0(e)$, a.u. & $\Delta\alpha(0)/\alpha_0(g)$\\ 
\rule{0pt}{4pt}\\
\hline
\rule{0pt}{4pt}\\
72 & Hf$^{12+}$ & 0.266690 & 0.267220 & 0.00199\\
\rule{0pt}{2pt}\\
\hline
\rule{0pt}{2pt}\\
74 & W$^{14+}$ & 0.164300 & 0.164560 & 0.00158\\
\rule{0pt}{2pt}\\
\hline
\rule{0pt}{2pt}\\
76 & Os$^{18+}$ & 0.110040 & 0.110150 & 0.00127\\
\rule{0pt}{2pt}\\
\hline
\rule{0pt}{2pt}\\
78 & Pt$^{20+}$ & 0.081409 & 0.081482 & 0.00090\\
\rule{0pt}{2pt}\\
\hline
\rule{0pt}{2pt}\\
80 & Hg$^{22+}$ & 0.062654 & 0.062703 & 0.00078\\
\rule{0pt}{2pt}\\
\hline
\rule{0pt}{2pt}\\
82 & Pb$^{24+}$ & 0.049640 & 0.049675 & 0.00071\\
\rule{0pt}{2pt}\\
\hline
\rule{0pt}{2pt}\\
84 & Po$^{26+}$ & 0.040200 & 0.040225 & 0.00062\\
\rule{0pt}{2pt}\\
\hline
\rule{0pt}{2pt}\\
88 & Ra$^{30+}$ & 0.027645 & 0.027660 & 0.00054\\
\rule{0pt}{2pt}\\
\hline
\rule{0pt}{2pt}\\
90 & Th$^{32+}$ & 0.023338 & 0.023349 & 0.00047\\
\rule{0pt}{2pt}\\
\hline
\rule{0pt}{2pt}\\
92 & U$^{34+}$ & 0.019886 & 0.019895 & 0.00045\\
\rule{0pt}{2pt}\\
\hline\hline
\end{tabular}}\label{Tab:3}
\end{table}

\section{Scalar polarizability of different levels of the same configuration.}\label{Cancel}
Our numerical calculations of the polarizabilities have been performed using exact equation (\ref{alpha}).  In order to show that the scalar static polarizability has close values
for levels of the same configuration it is convenient
 to replace summation over exact eigenstates in
equation (\ref{alpha}) by the summation over single-particle excitations from the ground state:
\begin{equation}\label{alpha_i}
\alpha_0(a) = \frac{2}{3(2J_a+1)} \sum_b \frac{\sum_i\langle a||\mathbf{d}_i|| b
  \rangle^2}{E_a - E_b},
\end{equation}
where $\mathbf{d}_i$ is a single electron dipole moment operator. Lets
consider in details reduced matrix elements $\langle a||\mathbf{d}_i||
b \rangle$. It is convenient to expand wavefunctions of the system in
terms of non-relativistic configurations, so that apart from total
angular momentum, the state is described by total orbital momentum $L$
and total spin $S$. For simplicity lets consider two valence
electron system. In total sum (\ref{alpha_i}) lets separate contributions that correspond to electric dipole transition (E1) of
a single electron $n_1l_1$ to excited state $n_1'l_1'$. In this case
matrix elements in (\ref{alpha_i}) can be written as  
\begin{equation}\label{mat_elem}
 \langle a||\mathbf{d}_i|| b\rangle=\langle
 n_1l_1n_2l_2LSJ||\mathbf{d}_1||n_1'l_1'n_2l_2L'SJ'\rangle, 
\end{equation}
where operator $\mathbf{d}_1$ acts on 1-st electron with orbital
momentum $l_1$. To simplify the above expression it is convenient to
use formula (13.2.5) from \cite{Varshalovich} 
\begin{eqnarray}\label{step1}
&&\langle n_1l_1n_2l_2LSJ||\mathbf{d}_1||n_1'l_1'n_2l_2L'SJ'\rangle=(-1)^{J'+L+S+1}\times\nonumber\\
&& \Pi_{JJ'}\left\{ \begin{array}{lll} L' & S & L \\ J & 1 &
      J' \end{array} \right\}\langle n_1l_1n_2l_2L||\mathbf{d}_1||n_1'l_1'n_2l_2L'\rangle,
\end{eqnarray}
where $ \Pi_{JJ'}=\sqrt{(2J+1)(2J'+1)}$. Applying the same formula (13.2.5) for the orbital momentum part of wavefunction one can obtain the following expression
\begin{eqnarray}\label{step2}
&&\langle n_1l_1n_2l_2L||\mathbf{d}_1||n_1'l_1'n_2l_2L'\rangle=(-1)^{L'+l_1+l_2+1}\times\nonumber\\
&& \Pi_{LL'}\left\{ \begin{array}{lll} l_1' & l_2 & l_1 \\ L & 1 &
      L' \end{array} \right\}\langle n_1l_1||\mathbf{d}_1||n_1'l_1'\rangle.
\end{eqnarray} 
Substituting (\ref{step1}), (\ref{step2}) in (\ref{mat_elem}) one can obtain the following relation
\begin{eqnarray}\label{sum_elem}
&&\frac{2}{3\Pi_J^2}\sum_{L',J'}\langle n_1l_1n_2l_2LSJ||\mathbf{d}_1||n_1'l_1'n_2l_2L'SJ'\rangle^2=\nonumber\\
&&\frac{2}{3}\sum_{L',J'}\Pi_{J',L',L}^2\left\{\begin{array}{lll} l_1' & l_2 & l_1 \\ L & 1 &
      L' \end{array} \right\}^2\left\{ \begin{array}{lll} L' & S & L \\ J & 1 &
      J' \end{array} \right\}^2\times\nonumber\\
&&\langle n_1l_1||\mathbf{d}_1||n_1'l_1'\rangle^2.
\end{eqnarray}
Using formula (12.2.7) from Ref.~\cite{Varshalovich} to carry out summation over $J'$ in the above equation one gets
\begin{eqnarray}\label{sum_elem_1}
&&\frac{2}{3\Pi_J^2}\sum_{L',J'}\langle n_1l_1n_2l_2LSJ||\mathbf{d}_1||n_1'l_1'n_2l_2L'SJ'\rangle^2=\nonumber\\
&&\frac{2}{3}\langle n_1l_1||\mathbf{d}_1||n_1'l_1'\rangle^2\sum_{L'}\Pi_L'^2\left\{ \begin{array}{lll} l_1' & l_2 & l_1 \\ L & 1 &
      L' \end{array} \right\}^2,
\end{eqnarray}
and employing the same formula (12.2.7) again to sum over $L'$, one obtains the following equation
\begin{eqnarray}\label{sum_elem_2}
&&\frac{2}{3\Pi_J^2}\sum_{L',J'}\langle n_1l_1n_2l_2LSJ||\mathbf{d}_1||n_1'l_1'n_2l_2L'SJ'\rangle^2=\nonumber\\
&&\frac{2}{3\Pi_{l_1}}\langle n_1l_1||\mathbf{d}_1||n_1'l_1'\rangle^2.
\end{eqnarray}

Summation in the above equation is over orbital momentum $L'$ and
total angular momentum $J'$. Energy levels
$|n_1'l_1'n_2l_2L'SJ'\rangle$ are assumed to be degenerate over these
quantum numbers. It follows from (\ref{sum_elem_2}) that (\ref{alpha_i})
doesn't depend on $L, J$ of state $|a\rangle$ but only on electron
configuration $|n_1l_1n_2l_2\rangle$. Similar property of scalar
static polarizabilities were obtained in \cite{Angel:1968} but using the
assumption that the basis set is completely degenerate. In real atoms the spin-orbit interaction
removes degeneracy  for states with different $J'$ of the
same $^{2S+1}L'$ multiplet. For different multiplets it is removed by
both the spin-orbit and the Coulomb interaction. This makes above
statement about the scalar polarizabilities of all states of the same
configuration to be independent on $L, J$ to hold only
approximately. In order to demonstrate this, the  accurate numerical
calculations of polarizabilities for tin ($5p^2$) and doubly ionized
zirconium ($4d^2$) were performed. Table \ref{Tab:4} presents
results of calculations performed in both relativistic (the fine structure
constant $\alpha=1/137$) and non-relativistic ($\alpha \rightarrow 0$)
formalisms. As one can see from table \ref{Tab:4}, the statement on
equality of the scalar polarizabilities for different states of the same
configuration is an approximation even in the  non-relativistic
approach. Although the non-relativistic solution returns exactly equal
scalar polarizabilities for all states of the same multiplet
 ,polarizabilities differ for different
multiplets. Indeed, absence of the $LS$ splitting leads to equality of
energy denominators within one multiplet, so the above conclusion can be
applied to the matrix elements in (\ref{alpha_i}). 
It is interesting to note very close values of the polarizabilities of
the states with the same total spin $S$.

Above situation significantly simplifies for highly charged ions
(HCI). It corresponds to the large spin-orbit interaction case, hence states of HCI
are well described in terms of $jj$ coupling. Table \ref{Tab:3}
represents results of calculations of the polarizabilities for HCI with
two holes in $4f$-shell. Difference of the polarizabilities between
selected components of $4f_{7/2}4f_{7/2}$ two hole states is several
orders of magnitude smaller compared to the values itself. Explanation
for this can be found in \cite{Kozlov:2014} and is similar to the explanation
presented above except that it is done in relativistic formalism. This
mechanism works well for the Pa$^{3+}$ ion considered in this work
(see Table~\ref{Tab:2}).  

\section{Results}

\begin{table*}\center
\caption{Clock transitions in neutral atoms and low charged ions with suppressed BBR shift.}
{\renewcommand{\arraystretch}{0}%
\begin{tabular}{l c c l l l l l l l}
\hline\hline
\rule{0pt}{4pt}\\
Z & element & clock states & term & calc. energy, & exp. energy, &  $\alpha_0(a)$, & $Q_a$, & $\Gamma$, & 1/Q\\
\rule{0pt}{2pt}\\
 & & & & cm$^{-1}$ & cm$^{-1}$~\cite{CRCHndbook, FrenchTxtbook} & a.u. & $|e|a_0^2$ & $\mu$Hz & \\
\rule{0pt}{4pt}\\
\hline
\rule{0pt}{2pt}\\
 \multirow{2}{*}{40} &  \multirow{2}{*}{Zr$^{2+}$} & $4d^2$ & $^3\text{F}_2$ & 0 & 0 & 11.05 & -0.89 &\multirow{2}{*}{$7853$} &\multirow{2}{*}{$5.2\times10^{-18}$}\\
\rule{0pt}{2pt}\\
 & & $4d^2$ & $^3\text{P}_0$ & 7902 & 8063 & 11.29 & 0 & & \\
\rule{0pt}{2pt}\\
\hline
\rule{0pt}{2pt}\\
 \multirow{2}{*}{40} &  \multirow{2}{*}{Zr} & $4d^25s^2$ & $^3\text{F}_2$ & 0 & 0 & 129.8 & -0.37 & \multirow{2}{*}{$18.7$} &\multirow{2}{*}{$2.4\times10^{-20}$}\\
\rule{0pt}{2pt}\\
 & & $4d^25s^2$ & $^3\text{P}_0$ & 4332 & 4196 & 138.6 & 0 & & \\
\rule{0pt}{2pt}\\
\hline
\rule{0pt}{2pt}\\
 \multirow{2}{*}{52} &  \multirow{2}{*}{Te} & $5p^4$ & $^3\text{P}_2$ & 0 & 0 & 45.96 & -2.58 & \multirow{2}{*}{$1745$} &\multirow{2}{*}{$2.0\times10^{-18}$}\\
\rule{0pt}{2pt}\\
 & & $5p^4$ & $^3\text{P}_0$ & 4736 & 4706 & 47.80 & 0 & & \\
\rule{0pt}{2pt}\\
\hline
\rule{0pt}{2pt}\\
 \multirow{2}{*}{53} &  \multirow{2}{*}{I$^{+}$} & $5p^4$ & $^3\text{P}_2$ & 0 & 0 & 22.08 & -1.64 & \multirow{2}{*}{$4279$} &\multirow{2}{*}{$3.5\times10^{-18}$}\\
\rule{0pt}{2pt}\\
 & & $5p^4$ & $^3\text{P}_0$ & 6643 & 6447 & 22.48 & 0 & & \\
\rule{0pt}{2pt}\\
\hline
\rule{0pt}{2pt}\\
 \multirow{2}{*}{54} &  \multirow{2}{*}{Xe$^{2+}$} & $5p^4$ & $^3\text{P}_2$ & 0 & 0 & 14.69 & -1.17 & \multirow{2}{*}{$8756$} &\multirow{2}{*}{$5.7\times10^{-18}$}\\
\rule{0pt}{2pt}\\
 & & $5p^4$ & $^3\text{P}_0$ & 8459 & 8130 & 14.79 & 0 & & \\
\rule{0pt}{2pt}\\

\hline
\rule{0pt}{2pt}\\
 \multirow{2}{*}{84} &  \multirow{2}{*}{Po} & $6p^4$ & $^3\text{P}_2$ & 0 & 0 & 54.55 & -1.34& \multirow{2}{*}{$35709$} &\multirow{2}{*}{$2.5\times10^{-17}$}\\
\rule{0pt}{2pt}\\
 & & $6p^4$ & $^3\text{P}_0$ & 7989 & 7514 & 59.41 & 0 & & \\
\rule{0pt}{2pt}\\

\hline
\rule{0pt}{2pt}\\
 \multirow{2}{*}{72} &  \multirow{2}{*}{Hf} & $5d^26s^2$ & $^3\text{F}_2$ & 0 & 0 & 102.4 & -0.84 & \multirow{2}{*}{$668$} &\multirow{2}{*}{$6.4\times10^{-19}$}\\
\rule{0pt}{2pt}\\
 & & $5d^26s^2$ & $^3\text{P}_0$ & 5172 & 5521 & 118.9 & 0 & & \\
\rule{0pt}{2pt}\\
\hline
\rule{0pt}{2pt}\\
 \multirow{2}{*}{90} &  \multirow{2}{*}{Th} & $6d^27s^2$ & $^3\text{F}_2$ & 0 & 0 & 163& -1.23 & \multirow{2}{*}{$22.3$} &\multirow{2}{*}{$4.6\times10^{-20}$}\\
\rule{0pt}{2pt}\\
 & & $6d^27s^2$ & $^3\text{P}_0$ & 2187 & 2558 & 165.7 & 0 & & \\
\rule{0pt}{2pt}\\
\hline
\rule{0pt}{2pt}\\
 \multirow{2}{*}{68} &  \multirow{2}{*}{Er} & $4f^{12}6s^2$ & $^3\text{H}_6$ & 0 & 0 & 150.2 & 0.71 & \multirow{2}{*}{$25.1$} &\multirow{2}{*}{$2.6\times10^{-20}$}\\
\rule{0pt}{2pt}\\
 & & $4f^{12}6s^2$ & $^3\text{F}_4$ & 6169 & 5035 & 150.2 & -0.01 & & \\
\rule{0pt}{2pt}\\

\hline
\rule{0pt}{2pt}\\
 \multirow{2}{*}{91} &  \multirow{2}{*}{Pa$^{3+}$} & $5f^2$ & $^3\text{H}_4$ & 0 & 0 & 9.86 & -1.12 & \multirow{2}{*}{$19.6$} &\multirow{2}{*}{$3.6\times10^{-20}$}\\
\rule{0pt}{2pt}\\
 & & $5f^2$ & $^3\text{F}_2$ & 3329 & 2878 & 10.07 & 0.30 & & \\
\rule{0pt}{2pt}\\
\hline
\rule{0pt}{2pt}\\
 \multirow{2}{*}{91} &  \multirow{2}{*}{Pa$^{3+}$} & $5f^2$ & $^3\text{F}_2$ & 3329 & 2878 & 10.07 & 0.30 &\multirow{2}{*}{$3467$} &\multirow{2}{*}{$2.1\times10^{-18}$}\\
\rule{0pt}{2pt}\\
 & & $5f^2$ & $^3\text{P}_0$ & 12989 & 11512 & 10.67 & 0 & & \\
\rule{0pt}{2pt}\\
\hline
\rule{0pt}{2pt}\\
 \multirow{2}{*}{68} &  \multirow{2}{*}{Er$^{2+}$} & $4f^{12}$ & $^3\text{H}_6$ & 0 & 0 & 3.91 & 0.40 & \multirow{2}{*}{$8.4$} &\multirow{2}{*}{$8.7\times10^{-21}$}\\
\rule{0pt}{2pt}\\
 & & $4f^{12}$ & $^3\text{F}_4$ & 6159 & 5081 & 2.80 & -0.02& & \\
\rule{0pt}{2pt}\\
\hline
\rule{0pt}{2pt}\\
 \multirow{2}{*}{69} &  \multirow{2}{*}{Tm$^{3+}$} & $4f^{12}$ & $^3\text{H}_6$ & 0 & 0 & 0.85 & 0.30 & \multirow{2}{*}{$8.6$} &\multirow{2}{*}{$8.1\times10^{-21}$}\\
\rule{0pt}{2pt}\\
 & & $4f^{12}$ & $^3\text{F}_4$ & 6714 & 5640 & 0.80 & -0.01 & & \\
\rule{0pt}{2pt}\\
\hline
\rule{0pt}{2pt}\\
 \multirow{2}{*}{71} &  \multirow{2}{*}{Lu$^{+}$} & $6s^{2}$ & $^1\text{S}_0$ & 0 & 0 & 63.10 & 0 & \multirow{2}{*}{$12.2$} &\multirow{2}{*}{$4.2\times10^{-21}$}\\
\rule{0pt}{2pt}\\
 & & $5d 6s$ & $^3\text{D}_1$ & 11995 & 11796 & 60.87 & 0 & & \\
\rule{0pt}{4pt}\\
\hline\hline
\end{tabular}}\label{Tab:2}
\end{table*}

List of suitable elements for application in ion clock and optical
lattice atomic clocks is presented in Table~\ref{Tab:2}. The rest of
the neutral or low charged ions with two electrons or holes in opened
shell have either no suitable clock transition or have Q-factors under
$10^{17}$. Values of the fractional BBR shifts of the clock transition
frequency at room temperature are calculated using Eq. (\ref{BBR}) and are
presented in Table~\ref{Tab:1}. It shows that most of the considered
neutral elements have BBR shift at room temperature of the same order
as Ag clock~\cite{Topcu:2006}. Calculations were performed using the configuration
interaction (CI) and the many body perturbation theory (MBPT)
method. Detailed description of the method can be found in our recent
papers~\cite{Dzuba:2005, Kozlov:2014}. For the $4f^{12}$ opened shell
configuration like in Er, Er$^{2+}$ and Tm$^{3+}$ we use CI
calculations without MBPT \cite{Er}. The values of BBR
shift at room temperature for Er, Er$^{2+}$ and Tm$^{3+}$ are expected
to be overestimated due to the  low accuracy of employed CI method for 12
and 14 valence electrons and should be considered as an upper limit. 

Calculations of the quadrupole moments presented in Table \ref{Tab:2} show
that a quadrupole moment of an atomic ground state can have both positive
and negative sign. Indeed, sign of quadrupole moment originates from
reduced matrix element in (\ref{quad}), which   includes
angular and radial integration. Although the radial part of the
integral is always positive and proportional to the average squared radius
of an atom, the angular integral can have both signs. Another important
consequence of that explains relatively low quadrupole moments of Er,
Er$^{2+}$ and Tm$^{3+}$ compared to the rest of the elements. Those
elements acquire their quadrupole momenta due to presence of two holes
in 4$f$-shell. Average squared radius of 4$f$-shell is significantly
smaller than the ones of 6$s$, 6$p$, 5$d$-shells and is of the same
order of magnitude as the one of 4$d$-shell. Indeed, the  quadrupole moment
of zirconium ground state is only two times smaller than the one of
erbium. Therefore,  extrapolating to  the rest of lanthanides and
actinides with configurations $(n-2)f^Nns^2$ or $(n-2)f^N$ one can
expect them to have relatively small quadrupole momenta. 

As it was pointed in the end of section \ref{QS}, the  clock states with either
total angular momentum of an atom $F=0, 1/2$ or total electronic angular
momentum $J=0, 1/2$ have no quadrupole shift. Selection of the
following isotopes $^{91}$Zr ($I=5/2$), $^{127}$I ($I=5/2$),
$^{131}$Xe ($I=3/2$), $^{231}$Pa ($I=3/2$) would result in emerging of
the hyperfine component of the ground state with $F=1/2$, while the first
excited state for considered neutral atoms and ions of these elements
will have electronic angular momentum $J=0$ (the second excited state for
$^{231}$Pa$^{3+}$). Therefore, it is possible to completely cancel
quadrupole shift for proposed clock transitions in $^{91}$Zr,
$^{91}$Zr$^+$, $^{127}$I$^+$, $^{131}$Xe$^{2+}$, and
$^3\text{F}_2\rightarrow ^3\text{P}_0$ in $^{231}$Pa$^{3+}$. 

It should be noted that presented elements were chosen only due to
the presence of a clock transition between different states of the same
configuration. This guarantees cancellation of BBR shift of no less
than one order of magnitude. However, the 98\% cancellation of BBR shift
in aluminum ion clock occurs between levels of different
configurations. Calculations for Lu$^+$ shows similar two orders of
magnitude cancellation for the strongly forbidden M1 transition. Therefore,
neutral atoms and low charged ions considered in this paper can only
be a part of the full list of elements suitable for ultra-precise
atomic clocks with suppressed BBR shift. 

\section{Discussion of accuracy}

\begin{table*}\center
\caption{Energies and transition amplitudes of odd levels that contribute to polarizability of neutral thorium [Rn]$6d^27s^2$ ground state. This table displays only several levels with energies under 20000 cm$^{-1}$.}
{\renewcommand{\arraystretch}{0}%
\begin{tabular}{c c c c c}
\hline\hline
\rule{0pt}{4pt}\\
leading & total & \multicolumn{2}{c}{energy, cm$^{-1}$} & transition  \\ 
\rule{0pt}{2pt}\\
configuration & momentum & \cite{FrenchTxtbook} & our calculation & amplitude, a.u.  \\ 
\rule{0pt}{4pt}\\
\hline
\rule{0pt}{4pt}\\
$5f6d7s^2$ & 2 & 8243 & 9671 & -0.0852 \\
\rule{0pt}{2pt}\\
$5f6d7s^2$ & 3 & 10526 & 12222 & 0.4618\\
\rule{0pt}{2pt}\\
$6d7s^27p$ & 2 & 10783 & 10452 & 0.1345\\
\rule{0pt}{2pt}\\
$5f6d7s^2$ & 3 & 11241 & 13664 & -0.3952\\
\rule{0pt}{2pt}\\
$6d7s^27p$ & 1 & 11877 & 13204 & 0.3928\\
\rule{0pt}{2pt}\\
$5f6d7s^2$ & 2 & 12114 & 15147 & 0.5077\\
\rule{0pt}{2pt}\\
$6d7s^27p$ & 3 & 13945 & 13875 & 0.5466\\
\rule{0pt}{2pt}\\
$6d7s^27p$ & 2 & 14032 & 15357 & 0.1334\\
\rule{0pt}{2pt}\\
$5f6d7s^2$ & 1 & 14243 & 17155 & 0.3326\\
\rule{0pt}{2pt}\\
$6d^27s7p$ & 2 & 14465 & 13647 & 0.5827\\
\rule{0pt}{2pt}\\
$5f6d^27s$ & 3 & 15618 & 14484 & 0.1534\\
\rule{0pt}{2pt}\\
$6d^27s7p$ & 1 & 15736 & 15944 & 0.9344\\
\rule{0pt}{2pt}\\
$6d7s^27p$ & 2 & 16217 & 17707& -0.1304\\
\rule{0pt}{2pt}\\
$6d^27s7p$ & 2 & 17224 & 16608 & -0.4105\\
\rule{0pt}{2pt}\\
$5f6d7s^2$ & 1 & 17354 & 17511 & 0.5506\\
\rule{0pt}{2pt}\\
$6d7s^27p$ & 3 & 17411 & 16260 & 0.1057\\
\rule{0pt}{2pt}\\
$5f6d7s^2$ & 2 & 17847 & 19116 & -0.1011\\
\rule{0pt}{2pt}\\
$6d7s^27p$ & 3 & 18069 & 18270 & -0.3416\\
\rule{0pt}{2pt}\\
$6d^27s7p$ & 1 & 18614 & 18271 & -0.2221\\
\rule{0pt}{2pt}\\
$6d^27s7p$ & 3 & 18930 & 19000 & -0.0687\\
\rule{0pt}{2pt}\\
$6d^27s7p$ & 3 & 19503 & 18638 & -0.4439\\
\rule{0pt}{2pt}\\
$6d^27s7p$ & 2 & 19516 & 19401 & 0.1840\\
\rule{0pt}{2pt}\\
$6d^27s7p$ & 1 & 19817 & 21020 & -0.7016\\
\rule{0pt}{2pt}\\
$6d^27s7p$ & 3 & 20423 & 21248 & 0.4502\\
\rule{0pt}{2pt}\\
$6d7s^27p$ & 1 & 20423 & 21796 & -0.0428\\
\rule{0pt}{2pt}\\
$6d^27s7p$ & 2 & 20522 & 19763 & -0.1900\\
\rule{0pt}{4pt}\\
\hline\hline
\end{tabular}}\label{Tab:5}
\end{table*}

For calculations of the polarizabilities we employed the CI+MBPT method
\cite{Dzuba:2005, Kozlov:2014} for all elements except Er$^{2+}$
and Tm$^{3+}$ for which CI for many-valence-electron systems was used
\cite{Er}. The accuracy of the CI+MBPT method depends on the
number of valence electrons. Better than 3\% accuracy can
be achieved for two valence electron systems while for four electrons
the uncertainty is larger and can reach 6\%. It should be mentioned
that in \cite{Kozlov:2014} the CI+MBPT method was employed for lanthanides
and actinides with up to 16 electrons in open shells. In this paper we
used the same approach for Er. It was possible due to
the separation of the $f$-shell valence electrons from the $s$-,$p$-
and $d$-shell ones and attributing them to the core (see
\cite{Kozlov:2014} for details). This allowed to reduce the many
electron problem to 2-3 valence electrons. Estimated accuracy of this
approach for calculation of polarizabilities of lanthanides and
actinides was 13\%. One may argue that this accuracy is not
sufficiently high to claim strong cancellation of the polarizability
values. However, since we consider similar states and perform
identical calculations for both states we expect strong cancellations
of  the uncertainties similar to  the cancellations of the
polarizabilities.   

The Zr, Hf, and Th atoms require separate consideration due to larger
number of valence electrons. Each of these atoms have four valence electrons 
and accurate treatment of the interactions between them leads  to a very 
large configuration interaction matrix which is beyond our present
computational capabilities. Presented results were obtained by using
smaller number of allowed excitations compared to other atoms. Such
cut of the CI basis set led to some reduction of accuracy. We estimate
it on the level of 6\% compared to 3\% accuracy for atoms with two or
three valence electrons.

Comparing our result for Zr, Hf and Th with the
ones in \cite{Doolen, CRCHndbook} one can notice good agreement for Zr
and Hf and some disagreement for Th. We have no explanation for this
disagreement. We stress that we perform very similar calculations for
all three atoms, have similar accuracy for the energies and expect similar
accuracy for the polarizabilities.
Table \ref{Tab:5} presents some results of
our calculations for energies and transition amplitudes for levels of
odd parity with $J=1,2,3$ in interval of up to 20000 cm$^{-1}$ that
contribute to the scalar polarizability of the thorium ground state. 

For Er$^{2+}$ and Tm$^{3+}$ ions, which have only $f$ valence
electrons, the calculations were performed using the many-electron
version of the CI method~\cite{Er} which
has accuracy of about 20$\%$. Note that absence of $s$ or $p$ valence
electrons leads to small values of the polarizabilities and small
difference between polarizabilities of the ground and clock
states. Therefore, high accuracy of the calculations is less important
for such systems. 

\begin{acknowledgements}
This work was funded by the Australian Research Council.  
\end{acknowledgements}

\end{document}